\documentclass{iau}

\usepackage[T1]{fontenc}
\usepackage{caption}
\usepackage{natbib}
\usepackage{orcidlink}

\usepackage{siunitx}
\DeclareSIUnit\parsec{pc}

\newcommand{\hi}{{\rm H}\,{\small\rm I}}

\begin{document}

\title{BIG-SPARC: The new SPARC database}

\author{
        Konstantin Haubner\orcidlink{0009-0007-7808-4653}$^{1,2}$,
        Federico Lelli\orcidlink{0000-0002-9024-9883}$^1$,
        Enrico Di Teodoro\orcidlink{0000-0003-4019-0673}$^2$,\\
        Francis Duey\orcidlink{0009-0003-1662-5179}$^3$,
        Stacy McGaugh\orcidlink{0000-0002-9762-0980}$^3$,
        James Schombert\orcidlink{0000-0003-2022-1911}$^4$,\\
        Kelley M. Hess\orcidlink{0000-0001-9662-9089}$^{5,6}$,
        and the Apertif Team
        }

\affiliation{
    $^1$INAF -- Arcetri Astrophysical Observatory, Largo Enrico Fermi 5, 50125 Firenze, Italy\\email: {\tt konstantin.haubner@inaf.it}\\
    $^2$Dipartimento di Fisica e Astronomia, Università degli Studi di Firenze,\\via G. Sansone 1, 50019 Sesto Fiorentino, Firenze, Italy\\
    $^3$Department of Astronomy, Case Western Reserve University, Cleveland, OH 44106, USA\\
    $^4$Department of Physics, University of Oregon, Eugene, OR 97403, USA\\
    $^5$Department of Space, Earth and Environment, Chalmers University of Technology, Onsala Space Observatory, 43992 Onsala, Sweden\\
    $^6$ASTRON, the Netherlands Institute for Radio Astronomy, Postbus 2, 7990 AA, Dwingeloo, The Netherlands
    }

\lefttitle{Haubner et al.}
\righttitle{BIG-SPARC: The new SPARC database}

\begin{abstract}
The Surface Photometry and Accurate Rotation Curves (SPARC) database has provided the community with mass models for 175 nearby galaxies, allowing different research teams to test different dark matter models, galaxy evolution models, and modified gravity theories. Extensive tests, however, are hampered by the somewhat heterogeneous nature of the \hi\ rotation curves and the limited sample size of SPARC. To overcome these limitations, we are working on BIG-SPARC, a new database that consists of about 4000 galaxies with \hi\ datacubes from public telescope archives (APERTIF, ASKAP, ATCA, GMRT, MeerKAT, VLA, and WSRT) and near infrared photometry from \textit{WISE}. For these galaxies, we will provide homogeneously derived \hi\ rotation curves, surface brightness profiles, and mass models. BIG-SPARC is expected to increase the size of its predecessor by a factor of more than 20. This is a necessary step to prepare for the additional order of magnitude increase in sample size expected from ongoing and future \hi\ surveys with the Square Kilometre Array (SKA) and its pathfinders.
\end{abstract}


\maketitle

\section{Introduction}

The study of rotation curves (RCs) has been central to establishing the missing mass problem in galactic systems, one of the biggest open questions in modern physics \citep{Rubin1978, Bosma1981, vanAlbada1985}. While these mass discrepancies are commonly explained via dark matter (DM) halos, the physical nature of this matter remains unclear. As a result, the study of RCs continues to provide some of the strongest constraints on the nature and properties of DM. In this context, \hi\ observations represent one of the best tools to study galaxy RCs for two main reasons. First, the \hi\ disks of galaxies often extend beyond their stellar disks, therefore probing deep into the DM-dominated regions. Second, compared to stellar and ionized gas disks, \hi\ disks are dynamically cold with velocity dispersions of ${\sim}\SI{10}{\kilo\meter\per\second}$, so the observed rotation velocity is a close proxy of the circular velocity tracing the gravitational potential. However, on the downside, resolved \hi\ studies require time-costly interferometric radio observations, which have been limiting the overall sample sizes of RC studies so far.

In this regard, a large step forward came from the SPARC database, which combines 175 RCs collected from the literature with near infrared (NIR) surface photometry from the \textit{Spitzer} space telescope \citep{Lelli2016}. Thanks to the limited effects of internal dust extinction and recent star formation, NIR photometry is the best proxy for the stellar mass distributions of galaxies (e.g., \citeauthor{Verheijen2001a} \citeyear{Verheijen2001a}). As a result, SPARC enabled new investigations of the baryonic Tully-Fisher relation (BTFR, \citeauthor{Lelli2016BTFR} \citeyear{Lelli2016BTFR}, \citeyear{Lelli2019BTFR}, \citeauthor{Desmond2019BTFR} \citeyear{Desmond2019BTFR}, \citeauthor{McGaugh2021BTFR} \citeyear{McGaugh2021BTFR}) and a measurement of the Hubble constant $H_0$ \citep{Schombert2020H0}. In addition, it was crucial for establishing two new scaling relations, the central density relation \citep{Lelli2016CDR} and the radial acceleration relation (RAR, \citeauthor{McGaugh2016RAR} \citeyear{McGaugh2016RAR}, \citeauthor{Lelli2017RAR} \citeyear{Lelli2017RAR}, \citeauthor{Li2018RAR} \citeyear{Li2018RAR}). In the context of $\Lambda$CDM cosmology, SPARC allowed to investigate the properties of DM halos \citep{Katz2017Halos, Katz2019Halos, Li2019aHalos, Li2019bHalos, Li2020Halos, Li2022aHalos, Li2022bHalos} and the related missing baryons problem \citep{Katz2018MissBar}, while beyond that, it enabled tests of alternative theories such as Emergent Gravity \citep{Lelli2017EG} and Modified Newtonian Dynamics \citep{Petersen2020, Chae2020EFE, Chae2021EFE, Chae2022EFE}. Furthermore, SPARC has also been used to investigate stellar mass-to-light ratios \citep{Schombert2019ML, Schombert2022ML}, maximum disk fits \citep{Starkman2018Fits}, and the pitfalls of Bayesian RC fits \citep{Li2021Fits}.

\begin{table}[t]
    \footnotesize
    \captionsetup{singlelinecheck=off, labelfont=bf, font=footnotesize, labelformat=simple}
    \caption{Sources from which the BIG-SPARC data cubes were collected.}
    \vspace{-.4cm}
    \begin{center}
    \label{TabSources}
    \begin{tabular}{lccc}
        \hline 
        Literature source & Telescope & \# of cubes & \# of unique galaxies\\ \hline
        APERTIF$^1$ & WSRT-APERTIF & 5954 & 1740\\
        WALLABY DR1$^2$ & ASKAP & 592 & 486\\
        WHISP$^3$ & WSRT & 272 & 360\\
        Atlas\textsuperscript{3D}$^{\,4}$ & WSRT & 142 & 289\\
        BLUEDISK$^5$ & WSRT & 49 & 175\\
        PPZoA$^6$ & WSRT & 317 & 153\\
        Hickson Compact Groups$^7$ & VLA & 25 & 83\\
        LVHIS$^8$ & ATCA & 79 & 72\\
        VIVA$^9$ & VLA & 42 & 56\\
        VGS$^{10}$ & WSRT & 65 & 43\\
        HALOGAS$^{11}$ & WSRT & 24 & 39\\
        PHANGS-VLA$^{12}$ & VLA & 23 & 39\\
        LITTLE THINGS$^{13}$ & VLA & 40 & 38\\
        EDGES$^{14}$ & VLA & 24 & 29\\
        EveryTHINGS$^{15}$ & VLA & 38 & 26\\
        Ursa Major Cluster$^{16}$ & WSRT & 42 & 26\\
        HIX$^{17}$ & ATCA & 23 & 25\\
        SAURON$^{18}$ & WSRT & 10 & 23\\
        LSB Galaxies$^{19}$ & VLA, WSRT & 32 & 22\\
        FIGGS$^{20}$ & GMRT & 22 & 22\\
        THINGS$^{21}$ & VLA & 33 & 22\\
        Starburst Dwarfs$^{22}$ & VLA, WSRT & 18 & 14\\
        MeerKAT Fornax Cluster$^{23}$ & MeerKAT & 2 & 3\\
        Individual studies & Arecibo, ATCA, VLA, WSRT & 46 & 97\\
        \textbf{Total} & & 7914 & 3882\\
        \hline
    \end{tabular}
    \end{center}
    \textbf{Notes:} $^1$\cite{Adams2022}; Hess et al. (in prep.), $^2$\cite{Koribalski2020, Westmeier2022, Deg2022}, $^3$\cite{Swaters2002}, $^4$\cite{Serra2012}, $^5$\cite{Wang2013}, $^6$\cite{Ramatsoku2016}, $^7$\cite{Jones2023}, $^8$\cite{Koribalski2018}, $^9$\cite{Chung2009}, $^{10}$\cite{Kreckel2012}, $^{11}$\cite{Heald2011}, $^{12}$\cite{Sun2020}, $^{13}$\cite{Hunter2012}, $^{14}$\cite{Richards2016, Richards2018}, $^{15}$\cite{Chiang2024}, $^{16}$\cite{Verheijen2001b}, $^{17}$\cite{Lutz2018}, $^{18}$\cite{Morganti2006}, $^{19}$\cite{vdHulst1993, deBlok1996, Pickering1997, Trachternach2009}, $^{20}$\cite{Begum2008}, $^{21}$\cite{Walter2008}, $^{22}$\cite{Lelli2014}, $^{23}$\cite{Serra2023}
\end{table}

\begin{figure}[t]
    \centering
    \includegraphics[width = \textwidth]{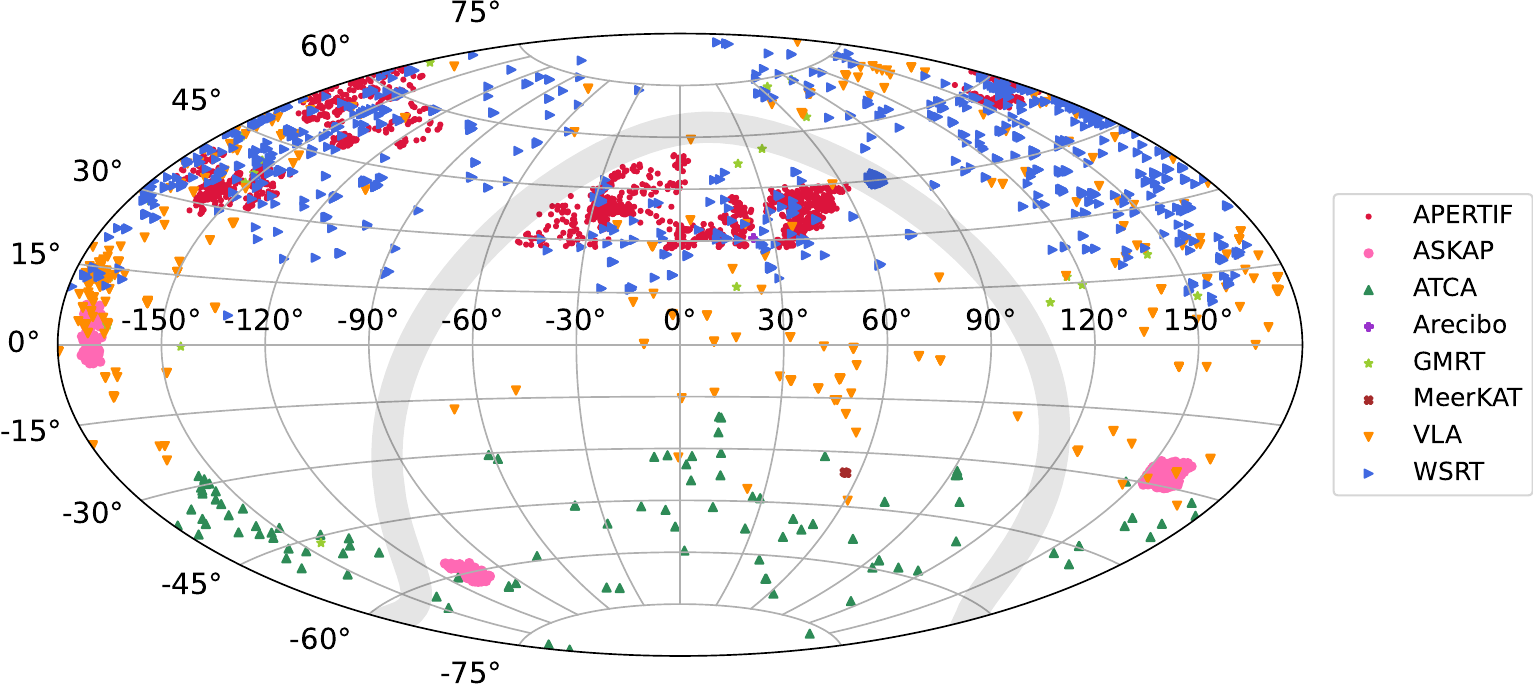}
    \caption{All-sky distribution of BIG-SPARC galaxies, color-coded by the telescopes that obtained their {\rm H}\,{\rm\fontsize{9pt}{10pt}\selectfont I}\ data. The gray band indicates the disk of the Milky Way.}
    \label{FigAllSky}
\end{figure}

\begin{figure}[h]
    \centering
    \includegraphics[width = \textwidth]{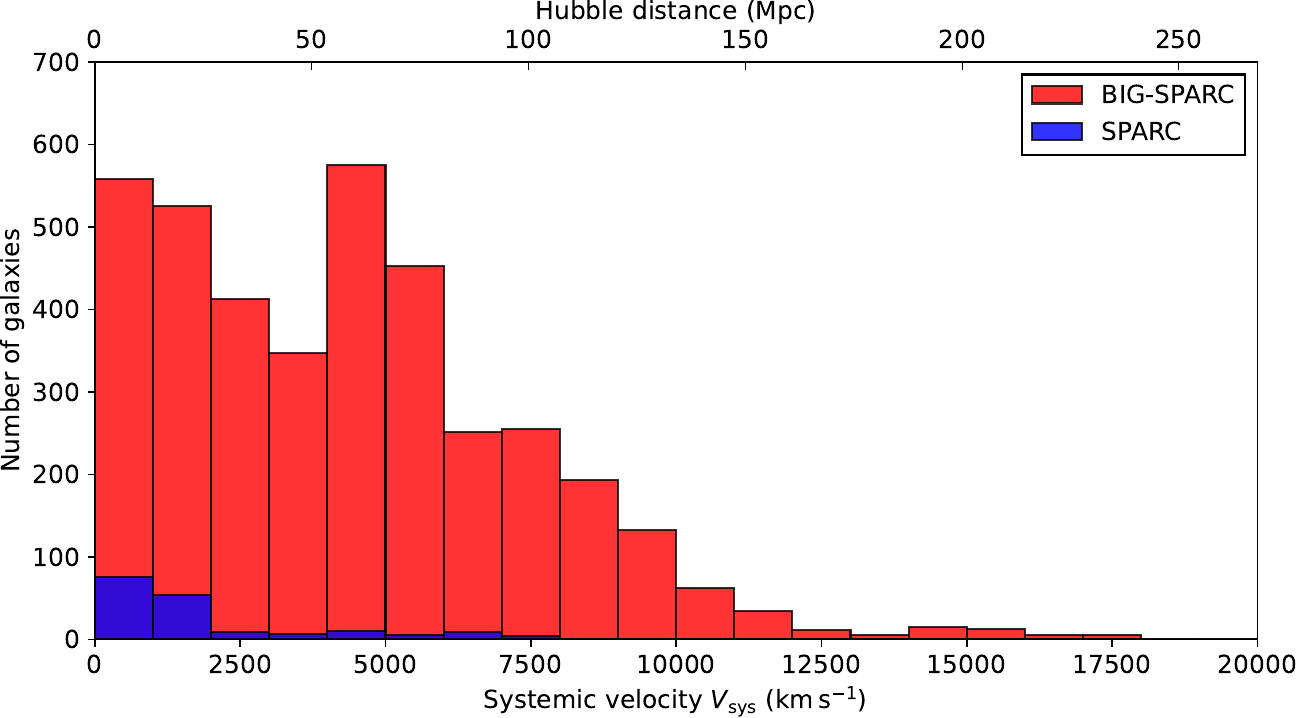} 
    \caption{Systemic velocities of BIG-SPARC galaxies (red histogram) compared to those of SPARC (blue histogram). The upper axis shows the corresponding Hubble distances.}
    \label{FigHubbleDistances}
\end{figure}

Despite its versatility, SPARC still suffers from two limitations. First, with 175 galaxies, the statistical power is limited. For example, at fixed galaxy mass, one cannot extensively study the effects of secondary properties such as environment, gas fraction, and so on.
Second, the RCs in SPARC were compiled from the literature, so some inhomogeneity unavoidably persists despite the undertaken homogenization efforts (see \citeauthor{Lelli2016} \citeyear{Lelli2016}). In particular, this limits the ability to study the very small intrinsic scatter around the scaling relations, which requires a perfect control of the observational uncertainties. To address these issues and to pave the way for the analysis of large future \hi\ surveys with the Square Kilometre Array (SKA) and its pathfinders, we are currently working on the successor of SPARC.

\section{Source finding and identification}

Differently from the previous SPARC database, we decided to derive RCs in a fully homogeneous way, starting from the \hi\ data cubes. Table~\ref{TabSources} lists the surveys and individual studies from which we culled our data. In total, we have 7914 cubes. To identify sources and produce \hi\ moment maps, we ran the source finding task \texttt{SEARCH} of the \textsuperscript{3D}Barolo software \citep{DiTeodoro2015} on all data cubes and crossmatched the resulting detections with the Principal Galaxies Catalogue (PGC, \citeauthor{Paturel2003} \citeyear{Paturel2003}). If this did not provide a match, we crossmatched with the NASA/IPAC Extragalactic Database (NED)\footnote{https://ned.ipac.caltech.edu/}. We then visually inspected all detections to remove artifacts. This procedure resulted in a list of almost 4000 galaxies. Due to the large sample size, we call the new data base the Broad Inventory of Galaxies with Surface Photometry and Accurate Rotation Curves (BIG-SPARC). The number of galaxies contributed by each survey or literature study is given in the last column of Table~\ref{TabSources}.

\begin{figure}[t]
    \centering
    \includegraphics[width = \textwidth]{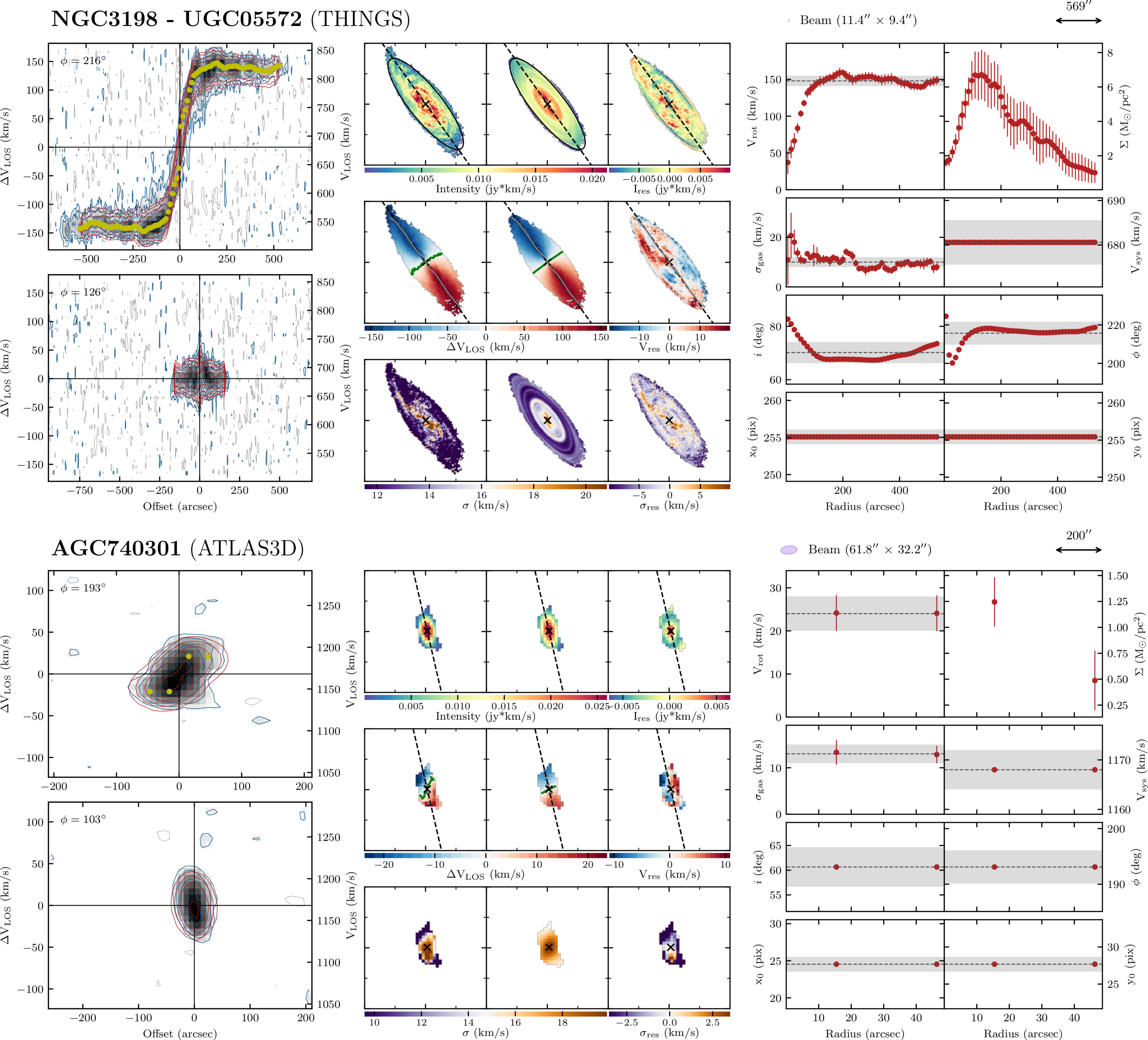} 
    \caption{Outputs of \texttt{3DFIT} for two BIG-SPARC galaxies: NGC~3198 (\textit{upper panel}, well resolved) and AGC~740301 (\textit{lower panel}, poorly resolved). The left column shows position-velocity diagrams along the galaxies' major and the minor axes. The middle column shows moment-0, -1, and -2 maps from observations (left), best-fit models (middle), and residuals (right). The right column shows best-fit parameters at different radii $R$ (from left to right and top to bottom): rotation velocity $V_\mathrm{rot}$, {\rm H}\,{\rm\fontsize{9pt}{10pt}\selectfont I}\ surface density $\Sigma$, velocity dispersion $\sigma_\mathrm{gas}$, systemic velocity $V_\mathrm{sys}$, inclination $i$, position angle $\phi$, and coordinates of the kinematic center $(x_0, y_0)$.}
    \label{FigKinematicFits}
\end{figure}

\section{All-sky distribution and distances}

Figure~\ref{FigAllSky} shows the all-sky distribution of the BIG-SPARC galaxies in equatorial coordinates, where the gray band indicates the disk of the Milky Way. The galaxies are color-coded based on the telescopes that obtained their \hi\ data. In general, the northern hemisphere is covered more densely than the southern one, which is mainly an effect of the Westerbork Synthesis Radio Telescope (WSRT) and its upgraded version APERTIF. Compared to WSRT, the Very Large Array (VLA) reaches lower declinations and connects with the coverage of the Australia Telescope Compact Array (ATCA) and the Australian SKA Pathfinder (ASKAP) in the south.

Figure~\ref{FigHubbleDistances} shows the systemic velocity histogram of the BIG-SPARC galaxies in red and that of the first SPARC database in blue. Compared to SPARC, BIG-SPARC provides an order-of-magnitude improvement in sample size. On the upper axis, we show the Hubble distances calculated with $H_0 = \SI{75}{\kilo\meter\per\second\per\mega\parsec}$ \citep{Tully2016}. While the farthest galaxy in SPARC, NGC~6195, has a distance of ${\sim}\SI{128}{\mega\parsec}$, BIG-SPARC reaches distances almost twice as large. The final BIG-SPARC database will contain direct distance measurements taken from the Cosmicflows-4 database \citep{Tully2023} where available and Hubble distances in the remaining cases.

\section{Outlook}

The kinematic fits to the BIG-SPARC galaxies will be performed with the \texttt{3DFIT} task of \textsuperscript{3D}Barolo. Two extreme examples are shown in Fig.~\ref{FigKinematicFits}: a well resolved spiral galaxy from THINGS (NGC~3198) and a poorly resolved dwarf galaxy from Atlas\textsuperscript{3D} (AGC~740301). The fit outputs highlight one of the main advantages of \textsuperscript{3D}Barolo, namely its ability to produce reasonable kinematic models for diverse data qualities. In addition, to allow the calculation of baryonic mass models, BIG-SPARC will contain NIR surface photometry wherever available. Since \textit{Spitzer} imaging is available for only a fraction of the BIG-SPARC galaxies, we will also exploit \textit{WISE} W1 photometry, which has full sky coverage (\citeauthor{Duey2024} \citeyear{Duey2024}, in prep.).

\small
\section*{Acknowledgments}

This work makes use of data from the Apertif system installed at the Westerbork Synthesis Radio Telescope owned by ASTRON. ASTRON, the Netherlands Institute for Radio Astronomy, is an institute of the Dutch Research Council ("De Nederlandse Organisatie voor Wetenschappelijk Onderzoek", NWO). KMH acknowledges the Spanish Prototype of an SRC (SPSRC) service and support funded by the Ministerio de Ciencia, Innovación y Universidades (MICIU), by the Junta de Andalucía, by the European Regional Development Funds (ERDF), and by the European Union NextGenerationEU/PRTR. The SPSRC acknowledges financial support from the Agencia Estatal de Investigación (AEI) through the "Center of Excellence Severo Ochoa" award to the Instituto de Astrofísica de Andalucía (IAA-CSIC) (SEV-2017-0709) and from the grant CEX2021-001131-S funded by MICIU/AEI/10.13039/501100011033. This scientific work uses data obtained from Inyarrimanha Ilgari Bundara / the Murchison Radio-astronomy Observatory. We acknowledge the Wajarri Yamaji People as the Traditional Owners and native title holders of the Observatory site. CSIRO's ASKAP radio telescope is part of the Australia Telescope National Facility (https://ror.org/05qajvd42). Operation of ASKAP is funded by the Australian Government with support from the National Collaborative Research Infrastructure Strategy. ASKAP uses the resources of the Pawsey Supercomputing Research Centre. Establishment of ASKAP, Inyarrimanha Ilgari Bundara, the CSIRO Murchison Radio-astronomy Observatory and the Pawsey Supercomputing Research Centre are initiatives of the Australian Government, with support from the Government of Western Australia and the Science and Industry Endowment Fund.

\bibliographystyle{aa}
\bibliography{references}

\end{document}